# A microscopically accurate model of partially ballistic nanoMOSFETs in saturation based on channel backscattering

G. Giusi[(a)], G. Iannaccone [(b)], F. Crupi [(a)]

[(a)]DEIS, University of Calabria, Via P. Bucci 41C, I-87036 Arcavacata di Rende (CS), Italy

ggiusi@deis.unical.it

[(b)]DIIEIT and SEED Center, University of Pisa, Italy

## Abstract

We propose a model for partially ballistic MOSFETs and for channel backscattering that is alternative to the well known Lundstrom model and is more accurate from the point of view of the actual energy distribution of carriers. The key point is that we do not use the concept of "virtual source". Our model differs from the Lundstrom model in two assumptions: i) the reflection coefficients from the top of the energy barrier to the drain and from top of the barrier to the source are approximately equal (whereas in the Lundstrom model the latter is zero), and ii) inelastic scattering is assumed through a ratio of the average velocity of forward-going carriers to that of backward-going carriers at the top of the barrier $k_v > 1$ (=1 in the Lundstrom model). We support our assumptions with 2D full band Monte Carlo (MC) simulations including quantum corrections in nMOSFETs. We show that our model allows to extract from the electrical characteristics a backscattering coefficient very close to that obtained from the solution of the Boltzmann transport equation, whereas the Lundstrom model overestimates backscattering by up to 40%.

# I. Introduction

Charge transport in nanoscale MOSFETs requires a physical description that does not use the concept of mobility. One would really need analytical device models directly usable for extracting transport parameters from experimental characteristics [1-10]. Among these, the simplest and most successful is the Lundstrom model [2], based on the Natori theory for ballistic transport [1], which relies on the concept of *backscattering*. In the Lundstrom model the transport in the channel is regulated by the elastic injection and reflection of thermally distributed carriers at the *virtual source* (Fig.1). In saturation, the backscattering coefficient is defined as the ratio $I^-/I^+$ between the source injected current $I^+$ and the backscattered current $I^-$. The strength of the model is that it provides just a number, the backscattering coefficient *r*, which includes all scattering mechanisms in the channel and that it is easily extracted from I-V and C-V characteristics [11-19]. Quasi-ballistic transistors have *r* close to zero, so that all the injected carriers reach the drain side providing maximum current drive. Technology developers and transistor designers must aim at devices with low *r* in order to enhance performance. In this sense the backscattering coefficient is a parameter which provides information about the scalability of a given technology (material and/or architecture). The picture of the Lundstrom model has been revolutionary because it moved attention from the drain side to the source side. However, the assumption of elastic transport has attracted criticisms [19-20] as well as the specific expression for the backscattering coefficient [5]. In this paper we propose a charge transport model that is alternative to the Lundstrom model and is more accurate from the point of view of the actual energy distribution of carriers.

The remainder of the paper is divided as stated in the following. In Section II we briefly recall the Lundstrom backscattering model. In Section III the proposed model is presented. In Section IV the backscattering calculated with our model is compared with the backscattering calculated with the Lundstrom model and with the true value extracted by two dimensional Monte Carlo device simulations. Finally, conclusions are drawn in Section V.

## II. The Lundstrom Model

The Lundstrom model (LM) picture is illustrated in Fig.1. The model is one dimensional (1D) along the channel direction and only one subband ($E_1$) is considered populated. The top of the source to channel energy barrier is called the "virtual source" (VS) because carriers are considered to be injected by the source reservoir which extends from the source contact to the VS. Carriers inside the source reservoir (Fermi Level $E_{FS}$) are injected from the VS into the channel and constitute $I^+$. The positive directed moments ($I^+$, $n^+$) at the VS are assumed equal to the ballistic case ( $I^+_{S,BL}, n^+_{S,BL}$)

$$I^+ \approx I^+_{S,BL} \tag{1}$$

$$n^+ \approx n^+_{S,BL} \tag{2}$$

The ballistic directed moments ($I^+_{S,BL}, n^+_{S,BL}$) at the VS are calculated using the Natori model for ballistic transport [1]

$$I^+_{S,BL} = qW \frac{N_{2D}}{2} v_{th} \Im_{1/2}(\eta) \tag{3}$$

$$n^+_{S,BL} = \frac{N_{2D}}{2} \Im_0(\eta) \tag{4}$$

with

$$\eta = \frac{E_{FS} - E_1}{kT} \qquad N_{2D} = kT \frac{m_{DOS}}{\pi\hbar^2} \qquad v_{th} = \sqrt{\frac{2kT}{\pi m_C}}$$

where $q$ is the electronic charge, $W$ the device width, $N_{2D}$ the effective two dimensional density of states, $v_{th}$ the unidirectional thermal velocity, $k$ the Boltzmann constant, $\hbar$ the reduced Planck constant, $T$ the absolute temperature, $m_{DOS}$ the density of states effective mass, $m_C$ the conduction effective mass, $\Im_j$ the Fermi-Dirac integral of order $j$ and $E_1$ is the energy of the populated sub-band. The average velocity of the positive source injected component $v^+$ is equal to the ballistic case, that is

$$v^+ = \frac{I^+}{qWn^+} \approx \frac{I^+_{S,BL}}{qWn^+_{S,BL}} = v^+_{S,BL} = v_{th}\frac{\Im_{1/2}(\eta)}{\Im_0(\eta)} \tag{5}$$

In saturation drain injection is suppressed and the current $I^-$ at the VS is only due to a fraction $r$ of the source injected current $I^+$. The backscattering occurs in a *critical layer* ($l$) and it is assumed elastic, that is the average velocity of transmitted carriers is equal to the average velocity of backscattered carriers ($v^+ \approx v^-$). From the knowledge of the current $I_D$ and of the charge density $Q$ at the VS, the backscattering $r$ can be calculated by solving the coupled equations

$$\begin{aligned} I_D &= I^+ - I^- = (1-r)I^+ \approx (1-r)I^+_{S,BL} = \\ &= (1-r)qW\frac{N_{2D}}{2}v_{th}\Im_{1/2}(\eta) \end{aligned} \tag{6}$$

$$\begin{aligned} Q &= qn = q\left(n^+ + n^-\right) = n^+\left(1 + r\frac{v^+}{v^-}\right) \approx n^+_{S,BL}(1+r) = \\ &= \frac{N_{2D}}{2}\Im_0(\eta)(1+r) \end{aligned} \tag{7}$$

where the assumption of elastic scattering ($v^+ \approx v^-$) has been used in Eq. 7. Equations 6-7 can be compacted in the form

$$I_D = \frac{1-r}{1+r}WQv^+_{S,BL} \tag{8}$$

where the term $B$=(1-$r$)/(1+$r$) is referred as the *ballistic ratio.*

## III. Proposed Model

In our picture (illustrated in Fig. 2), we do not use the VS concept, and we treat in a symmetric way backscattering of forward-going and backward-going electrons. Exploiting current continuity, the source- and drain-injected ballistic component can be traced back to the physical injection contact: $I^+_{S,BL}$ ($I^-_{D,BL}$) at the source (drain) are due to carriers injected at $x_{S,inj}$ ($x_{D,inj}$) and with energy higher than $E_{TOP}$. In the absence of scattering between $x_{S,inj}$ ($x_{D,inj}$) and $x_{max}$, $I^+$ ($I^-$) at $x_{max}$ would be equal to $I^+_{S,BL}$ ($I^-_{D,BL}$). However, in the presence of scattering, the current $I^+$ ($I^-$) is the sum of the transmitted fraction 1-$r_{SD}$ (1-$r_{DS}$) of $I^+_{S,BL}$ ($I^-_{D,BL}$) and of the backscattered component $r_{SD}$ ($r_{DS}$) of $I^-$ ($I^+$)

$$I^+ = (1 - r_{SD})I^+_{S,BL} + r_{SD}I^- \tag{9}$$

$$I^- = (1 - r_{DS})I^-_{D,BL} + r_{DS}I^+ \tag{10}$$

where $r_{SD}$ ($r_{DS}$) is a backscattering coefficient between $x_{S,inj}$ ($x_{D,inj}$) and $x_{max}$. In saturation ($V_{DS}$>>$kT/q$), injection from the drain contact is suppressed and, neglecting the scattering at the drain, we get $I^-_{D,BL} \approx 0$. Moreover, if $r_{SD}$≈$r_{DS}$=$r$ we obtain from Eqs. 9-10

$$I^+ + I^- \approx I^+_{S,BL} + I^-_{D,BL} \approx I^+_{S,BL} \tag{11}$$

The model is completed by the same approximation used in the LM for the charge density (Eq. 2): $n^+$ is assumed to be equal the concentration $n^+_{S,BL}$ of forward-going carriers we would have at $x_{max}$ in the case of ballistic transport (Eq. 2). We can provide a rough justification for such approximation, that will be confirmed ex post in section IV by detailed Monte Carlo simulations. If we divide Eq. (11) by $qWv^+$ we get

$$n^+ \approx n^+_{S,BL}\frac{v^+_{S,BL}/v^+}{1+r} \tag{12}$$

Obviously, in the case of ballistic transport the fraction is equal to 1. If scattering increases, carriers injected from $x_{S,inj}$ lose energy due to inelastic scattering reducing their average velocity so that $v^+ < v^+_{S,BL}$ and the numerator in the fraction of Eq. 16 increases. At the same time the denominator (1+$r$) increases too. To simplify the model, we assume that these two effects compensate one another so that $n^+ \approx n^+_{S,BL}$.

Equation 11 is different from Eq.1 used in the LM, since we include in the model the scattering between $x_{S,inj}$ and $x_{max}$. As a matter of fact Eq. 9 reduces to the Lundstrom assumption (Eq. 1) when $r_{SD} \approx 0$. Based on Eqs. 2 and 11 and on the Natori equations (3-4), the drain current ($I_D$) and the total charge density ($Q$) at $x_{max}$ (which we prefer not to call VS anymore since we abandon the VS concept) can be calculated as

$$\begin{aligned} I_D &= I^+ - I^- = \frac{1-r}{1+r}\left(I^+ + I^-\right) \approx \frac{1-r}{1+r} I^+_{S,BL} = \\ &= \frac{1-r}{1+r} qW \frac{N_{2D}}{2} v_{th} \Im_{1/2}(\eta) \end{aligned} \tag{13}$$

$$\begin{aligned} Q &= qn = q\left(n^+ + n^-\right) = n^+\left(1 + r\frac{v^+}{v^-}\right) \approx n^+_{S,BL}\left(1 + rk_v\right) = \\ &= \frac{N_{2D}}{2} \Im_0(\eta)\left(1 + rk_v\right) \end{aligned} \tag{14}$$

The ratio $k_v = v^+/v^-$ is not assumed 1 as in the LM but is extracted directly from MC simulations, so that we do not assume elastic scattering at around $x_{max}$. As stated in the next section, it is approximately equal to 1.35 according to [20]. The average velocity of source injected carriers is found to be equal to

$$v^+ = \frac{I^+}{qWn^+} \approx \frac{I^+_{S,BL}}{qW(1+r)n^+_{S,BL}} = \frac{v^+_{S,BL}}{1+r} \tag{15}$$

Finally, Eqs. 13-14 can be compacted as

$$I_D = \frac{1-r}{(1+r)\,(1+rk_v)} WQv^+_{S,BL} \tag{16}$$

where the term in fraction is the ballistic ratio, which differs for the term $1+rk_v$ at the denominator with respect to the LM (Eq. 8), thus implying that the backscattering $r$ calculated with our model is expected to be lower with respect to the backscattering calculated by the Lundstrom model.

## IV. Validation by Monte Carlo Simulations

In order to develop a comparative analysis between the Lundstrom model and the proposed model, 2D semi-classical quantum corrected simulations were performed with the full band Monte Carlo (MC) simulator "MoCa", which includes all relevant scattering mechanisms [21, 22]. The simulated device (Fig. 3) is a double gate nMOSFET with a very thin undoped silicon body ($t_{Si}$=1.5nm) [4]. Such a thin body is chosen in order to match the 1D transport and the one sub-band hypothesis of the Natori model. We make the common assumption that only the first band of the unprimed ladder is occupied so that $m_{DOS}=2m_t$, $m_C=m_t$ where $m_t=0.19m_0$ is the transverse mass and $m_0$ is the electron free mass, as confirmed by Schrodinger-Poisson simulations. In Fig. 4 Eqs. 9-10 are solved, with respect to $r_{SD}$ and $r_{DS}$, for each point $x$ inside the channel for a device with channel length L=20nm. $\Gamma_{D,BL}$ is assumed 0 (saturation) so that $r_{DS}\approx\Gamma^-/\Gamma^+$ and the source injection point $x_{S,inj}$ is taken at the source/channel junction ($x$=-L/2). $\Gamma^+_{S,BL}$ is calculated by taking the energy distribution of the positive directed current at $x_{S,inj}$ ($\Gamma^+_S$) integrated for energies higher than the barrier height between $x_{S,inj}$ and $x_{max}$. The hypothesis $r_{SD}\approx r_{DS}$ is verified in a point very close to $x_{max}$ so that the approximation of Eq. 11 holds. To verify the hypothesis of the proposed model (Eqs. 2-11) with respect to the hypothesis of the LM (Eqs. 1-2), Eqs. 2-11 and Eqs. 1-2 have been inverted with respect to $\eta$ from the knowledge of $\Gamma^+$, $\Gamma^-$, $n^+$. The result is plotted in Fig. 5 for different values of L. As can be noticed, Eqs. 2-11 (proposed model) yield very similar values of $\eta$, while Eqs. 1-2 (LM) do not. A further proof of our assumptions is shown in Fig. 6 where ballistic and non-ballistic simulations are compared (in the ballistic case the scattering is turned off only in the channel). Fig. 6 (top) shows that the sum $\Gamma^++\Gamma^-$ at $x_{max}$ is close to that of the non-ballistic case (error 1.2%) confirming the hypothesis of Eq. 11, while $\Gamma^+$ differs significantly (9.3%) from the ballistic case (Eq. 1). Moreover Fig. 6 (bottom) shows that the hypothesis for $n^+$ (Eq. 2) is well verified (4.2%). From Fig.6 (top) one can also note that $\Gamma^++\Gamma^-$ differs significantly from $I^+$ for the ballistic case. The explanation can be found in Fig. 7 where the potential energy profile ($E_P$) and the average total energy ($E_T$) in the ballistic case are shown. Ballistic carriers lose energy close to the drain where they have a sufficient energy to be backscattered and surmount the channel energy barrier giving a contribution to $I^-$. However, when scattering is turned on in the channel, carriers arrive at the drain side with a lower energy and backscattered carriers at the drain have lower chances to surmount the channel energy barrier [20]. In Fig. 8, $r$ (top) and the $v^+$ (bottom), calculated with the proposed model and with the Lundstrom model, are compared with the results extracted directly from MC simulations for different values of L, $V_G$, $V_D$. Moreover, the backscattering obtained with the LM using the true $k_v=v^+/v^-$ extracted directly from MC simulation is shown. It can be observed that the backscattering coefficient extracted with the LM differs significantly from $I^-/I^+$ calculated by MC simulation (40-50%), and that using the true $k_v$ is not sufficient to compensate the gap (20-30%). As can be noticed, the proposed model reproduces very well the MC results for both $r$ and $v^+$. We find that $k_v=v^+/v^-$ is a weak function of device geometry and bias and is approximately equal to 1.35 (according to [20]). This number can be used for experimental extraction.

Finally let us discuss the two main limitations of the proposed method. The first limitation is that the proposed model assumes that the point where $r_{SD}\approx r_{DS}$ (let's call it $x_{cross}$) is very close to $x_{max}$. As a matter of fact Eq. 11 holds only at $x_{cross}$, whereas the Natori model (Eqs. 3-4) is valid only at $x_{max}$. Indeed the carrier distribution can be approximated by a Fermi-Dirac expression also for $x_{S,inj}\leq x\leq x_{max}$ so that, neglecting the small potential variation, Eqs (3-4) can be applied in this region. This means that our model is approximately valid until $x_{cross}\leq x_{max}$. Fig. 9 shows $x_{max}$ and $x_{cross}$ for different bias and for different gate lengths. It shows that for low gate voltage $x_{cross}\leq x_{max}$ and our model works. As the gate voltage increases, $x_{cross}$ moves towards the drain. For higher gate voltage $x_{cross}$ is significantly different from $x_{max}$, the carrier distribution is very different from a Fermi-Dirac distribution, the voltage drop with respect to $x_{max}$ is high so that Eqs (3-4) and the proposed model cannot be applied. Anyway in the simulation conditions the model continues to work with a gate overdrive of 0.6V (threshold voltage is 0.4V). Moreover, for a gate voltage in the working range ($V_G$=1V), drain voltage and channel length variations do not influence significantly the relative position of $x_{max}$ and $x_{cross}$.

Another limitation of the model is due to the single subband approximation. Schrodinger-Poisson simulations, performed with ATLAS, have been used to evaluate the percentage of occupation of the first sub-band ($E_1$) as a function of the silicon thickness. Fig. 10 shows that when the silicon thickness is increased above 2 nm the occupation of higher energy bands starts to become important. We argue that this is not a problem of our assumption ($r_{SD} \approx r_{DS}$ or Eq. 11) but it is related to the underlying Natori model, so that a similar problem is shared with the Lundstrom model. A multi-band version of our equations, using for example the approach proposed in [15], could be used to overcome this limitation.

## V. Conclusion

In this paper we have proposed a charge transport model for partially ballistic nanoMOSFETs in saturation based on channel backscattering, which is alternative to the well known Lundstrom model. In our picture we remove the concept of virtual source and we assume that equilibrium distributed carriers are injected in the channel from a source and from a drain injection point so that forward-going and backward-going fluxes are treated in a symmetrical way. The main difference with respect to the Lundstrom model is that we include in the model the scattering between the source injection point and the virtual source, leading to the result that the sum of the forward-going and of the backward going fluxes at the virtual source is approximately equal to the sum of the source and drain injected ballistic components. Moreover, our model takes into account for inelastic scattering at the virtual source by an approximately constant ratio between the average velocities of forward-going and backward-going fluxes. We have shown that, through two dimensional full band Monte Carlo simulations with quantum corrections, our model represents a significant improvement in terms of accuracy with respect to the model proposed by Lundstrom (up to 40%), and succeeds in connecting the backscattering coefficient with its true value that can be extracted through particle-based Monte Carlo simulations.

## References

[1] K. Natori, “Ballistic metal-oxide-semiconductor field effect transistor,” *J. Appl. Phys.*, vol. 76, pp. 4879–4890, Oct. 1994.

[2] M. S. Lundstrom, “Elementary scattering theory of the Si MOSFET”, *IEEE Electron Device Lett.*, vol. 18, pp. 361–363, July 1997.

[3] M. S. Lundstrom and Z. Ren, “Essential physics of carrier transport in nanoscale MOSFET’s”, *IEEE Trans. Electron Devices*, vol. 49, pp.133–141, Jan. 2002.

[4] A. Rahman, M.S. Lundstrom, “A compact scattering model for the nanoscale double-gate MOSFET”, *IEEE Trans. Electron Devices*, vol. 49, pp. 481–489, Mar. 2002.

[5] E. Fuchs, P. Dollfus, G. Le Carval, S. Barraud, D. Villanueva, F. Salvetti, H. Jaouen, T. Skotnicki, “A New Backscattering Model Giving a Description of the Quasi-Ballistic Transport in Nano-MOSFET”, *IEEE Trans. Electron Devices*, Vol. 52, NO. 10, October 2005.

[6] G. Mugnaini, G. Iannaccone, “Physic-Based Compact Model of Nanoscale MOSFETs – Part I: Transition From Drift-Diffusion to Ballistic Transport”, *IEEE Trans. Electron Devices*, Vol. 52, NO. 8, August 2005.

[7] G. Mugnaini, G. Iannaccone, “Physic-Based Compact Model of Nanoscale MOSFETs – Part II: Effects of degeneracy on transport”, *IEEE Transactions Electron Devices*, Vol. 52**,** n. 8, pp.1802-1806 (2005).

[8] M. J. Chen, L. F. Lu, “A Parabolic Potential Barrier-Oriented Compact Model for the $k_BT$ Layer’s Width in Nano-MOSFETs”, *IEEE Trans. Electron Devices*, Vol. 55, NO. 5, MAY 2008.

[9] G. Giusi, G. Iannaccone, M. Mohamed, U. Ravaioli, “Study of Warm Electron Injection in Double Gate SONOS by Full Band Monte Carlo Simulation”, *Electron Device Letters*, Vol. 29, Issue 11, pp. 1242 – 1244, 2008.

[10] J.-L.P.J. van der Steen, P. Palestri, D. Esseni and R.J.E. Hueting, “A New Model for the Backscatter Coefficient in Nanoscale MOSFETs”, *ESSDERC*, pp. 234-237, 2010

[11] M.J. Chen, H.-T. Huang, K.-C. Huang, P.-N. Chen, C.-S. Chang, and C.H. Diaz, "Temperature dependent channel backscattering coefficients in nanoscale MOSFETs", in *IEDM Tech. Dig.*, Dec. 2002, pp. 39–42.

[12] A. Lochtefeld , D. A. Antoniadis, "On experimental determination of carrier velocity in deeply scaled nMOS: how close to the thermal limit?", *IEEE Electron Dev Lett.*, 2001;22(2).

[13] A. Lochtefeld, I. J. Djomehri, G. Samudra and D.A. Antoniadis, "New insights into carrier transport in n-MOSFETs", *IBM J. Res. & Dev.*, 46, 347-357 (2002).

[14] A. Dobbie, B. De Jaeger, M. Meuris, T.E. Whall, E.H.C. Parker and D.R. Leadley, "Channel Backscattering Characteristics of High Performance Germanium pMOSFETs", *ULIS 2008*, pp. 7-10.

[15] V. Barral, T. Poiroux, J. Saint-Martin, D. Munteanu, J. L. Autran, and S. Deleonibus, "Experimental Investigation on the Quasi-Ballistic Transport: Part I-Determination of a New Backscattering Coefficient Extraction Methodology", *IEEE Trans. Electron Devices*, Vol. 56, NO. 3, March 2009.

[16] M. Zilli, P. Palestri, D. Esseni, L. Selmi, "On the experimental determination of channel back-scattering in nanoMOSFETs", *International Electron Device Meeting (IEDM) 2007*, pp. 105-108.

[17] G. Giusi, F. Crupi, P. Magnone, "Criticisms on and comparison of experimental channel backscattering extraction methods", Microelectronics Engineering, 88 (2011) 76–81, DOI: 10.1016/j.mee.2010.08.024.

[18] G. Giusi, G. Iannaccone, F. Crupi, "Barrier Lowering and Backscattering Extractionin Short-Channel MOSFETs", *IEEE Trans. Electron Devices*, Sep. 2010, Vol. 57, no.9, pp. 2132-2137.

[19] M. J. Chen, H. T. Huang, Y. C. Chou, R. T. Chen, Y. T. Tseng, P. N. Chen, C. H. Diaz, "Separation of Channel Backscattering Coefficients in Nanoscale MOSFETs", *IEEE Trans. Electron Devices*, Vol. 51, No.9, Sep. 2004.

[20] P. Palestri, D. Esseni, S. Eminente, C. Fiegna, E. Sangiorgi, L. Selmi," Understanding Quasi-Ballistic Transport in Nano-MOSFETs: Part I—Scattering in the Channel and in the Drain", *IEEE Trans. Electron Devices*, Vol. 52, No.12, pp. 2727-2735, Dec. 2005.

[21] G.A. Kathawala, B. Winstead, and U. Ravaioli, “Monte Carlo simulation of double-gate MOSFETs,” *IEEE Trans. Electron Devices*, Dec. 2003, Vol. 50, pp. 2467-2473.

[22] B. Winstead, and U. Ravaioli, “A Quantum Correction Based on Schrodinger Equation Applied to Monte Carlo Device Simulation”, *IEEE Trans. Electron Devices*, Feb. 2003, Vol. 50, no.2, pp. 440-446.

# Figure Caption

Figure 1

Lundstrom model picture. The model is 1D and only one subband ($E_1$) is considered populated. The top of the source to channel energy barrier is called the "virtual source" (VS) because carriers are considered to be injected by the source reservoir which extends from the source contact to the VS. Carriers inside the source reservoir (Fermi Level $E_{FS}$) are injected from the VS to the channel and constitute $I^+$. In saturation, a fraction of them ($r$) backscatters due to the scattering inside the critical layer ($l$) and constitutes $I^-$. The scattering is assumed elastic ($v^+ \approx v^-$) and the positive directed moments ($I^+$, $n^+$) are assumed to be equal to the ballistic case (Eqs. 1-4).

Figure 2

The proposed model picture. Carriers are injected from two injection points $x_{S,inj}$ and $x_{D,inj}$ by source ($I^+_S$) and drain ($I^-_D$) reservoirs into the channel. Their ballistic components ($I^+_{S,BL}$ and $I^-_{D,BL}$) will experience scattering going towards $x_{max}$. $I^+$ ($I^-$) is the positive (negative) directed current at $x_{max}$. Only a fraction 1-$r_{SD}$ (1-$r_{DS}$) of $I^+_{S,BL}$ ($I^-_{D,BL}$) will be a part of $I^+$ ($I^-$) and the rest will be backscattered towards the source (drain). The current $I^+$ ($I^-$) is completed by the backscattered component of $I^-$ ($I^+$) through the coefficient $r_{SD}$ ($r_{DS}$) (Eqs. 9-10).

Figure 3

The simulated structure is a DG nMOSFET with ultra-thin undoped silicon body ($t_{Si}$=1.5nm), oxide thickness $t_{ox}$=1.5 nm and long source/drain extensions ($L_{ext}$=35nm). The threshold voltage is 0.4V.

Figure 4

Backscattering along the channel. Eqs. 9-10 are solved with respect to $r_{SD}$ and $r_{DS}$, for each point $x$ inside the channel. $\Gamma_{D,BL}$ is assumed 0 (saturation) so that $r_{DS}=\Gamma^{-}/\Gamma^{+}$, and the source injection point ($x_{S,inj}$) is taken at the source/channel junction ($x$=-10nm). The hypothesis $r_{SD}\approx r_{DS}$ is verified very close to $x_{max}$ so that Eq. 11 holds.

Figure 5

Verification of the Lundstrom model and of the proposed model. Eqs. 1-2 and Eqs. 2-11 have been inverted in order to calculate $\eta$. Eqs. 2-11 (proposed model) give very similar values of $\eta$, while Eqs. 1-2 (Lundstrom) give different values.

Figure 6

Directed currents (top) and carrier density (bottom) along the device in the ballistic and not ballistic case. It is evident that assumption of Eq. 11 (proposed model) is better verified than assumption of Eq. 1 (Lundstrom model) (1.2% versus 9.3%). Moreover, the assumption of Eq. 2 (Lundstrom and proposed model) is well verified (4.3%)

Figure 7

Ballistic simulation of the potential energy profile ($E_P$), average total energy ($E_T$) and average total energy of positive and negative directed fluxes ($E^+$ and $E^-$). $E_T$ is at the level of $E_{TOP}$ at the drain junction so that the backscattering of the ballistic source injected carriers contributes to the negative flux in $x_{max}$.

Figure 8

Backscattering (a-c) and the positive directed velocity (d-f) calculated with the proposed model and with the Lundstrom model compared with the true results extracted directly from MC simulation for different gate lengths (left), gate voltages (middle) and drain voltages (right). The proposed model reproduces very well the MC results while the Lundstrom model overestimates them (40-50%). Moreover the backscattering obtained with the Lundstrom model using the true $k_v=v^+/v^-$ extracted directly from MC simulation is shown. It is found that it differs significantly from $I^-/I^+$ so that the approximation on $k_v$ is not sufficient to justify this gap. The negative directed velocity is also shown. It is found that $k_v=v^+/v^-$ is approximately equal to 1.35 according to [20].

Figure 9

Distance (in nm) between the source edge ($x_{S,inj}$) and $x_{max}$ (filled symbols) and $x_{cross}$ (empty symbols) for different bias and gate lengths. For low gate voltage we have $x_{cross} \leq x_{max}$ and the carrier distribution is close to a Fermi-Dirac, the Natori equations (3-4) are approximately valid, and the proposed model can be used. For higher gate voltages, $x_{cross}$ moves towards the drain where the carrier distribution cannot be approximated by the Fermi-Dirac so that Eqs (3-4) and the proposed model are no more valid. Moreover, for a gate voltage in the operating range ($V_G$=1V), drain voltage and channel length variations do not influence significantly the relative position of $x_{max}$ and $x_{cross}$.

Figure 10

First subband occupation as a function of silicon thickness, evaluated by Schrodinger-Poisson simulations performed with ATLAS. When the silicon thickness is increased above 2 nm the occupation of higher energy bands starts to become important. This is not a problem of our assumption ($r_{SD} \approx r_{DS}$ or Eq. 11) but it is related to the underlying Natori model, so that a similar problem is shared with the Lundstrom model. A multi-band version of our equations, using for example the approach proposed in [15], could be used to overcome this limitation.

# Figures

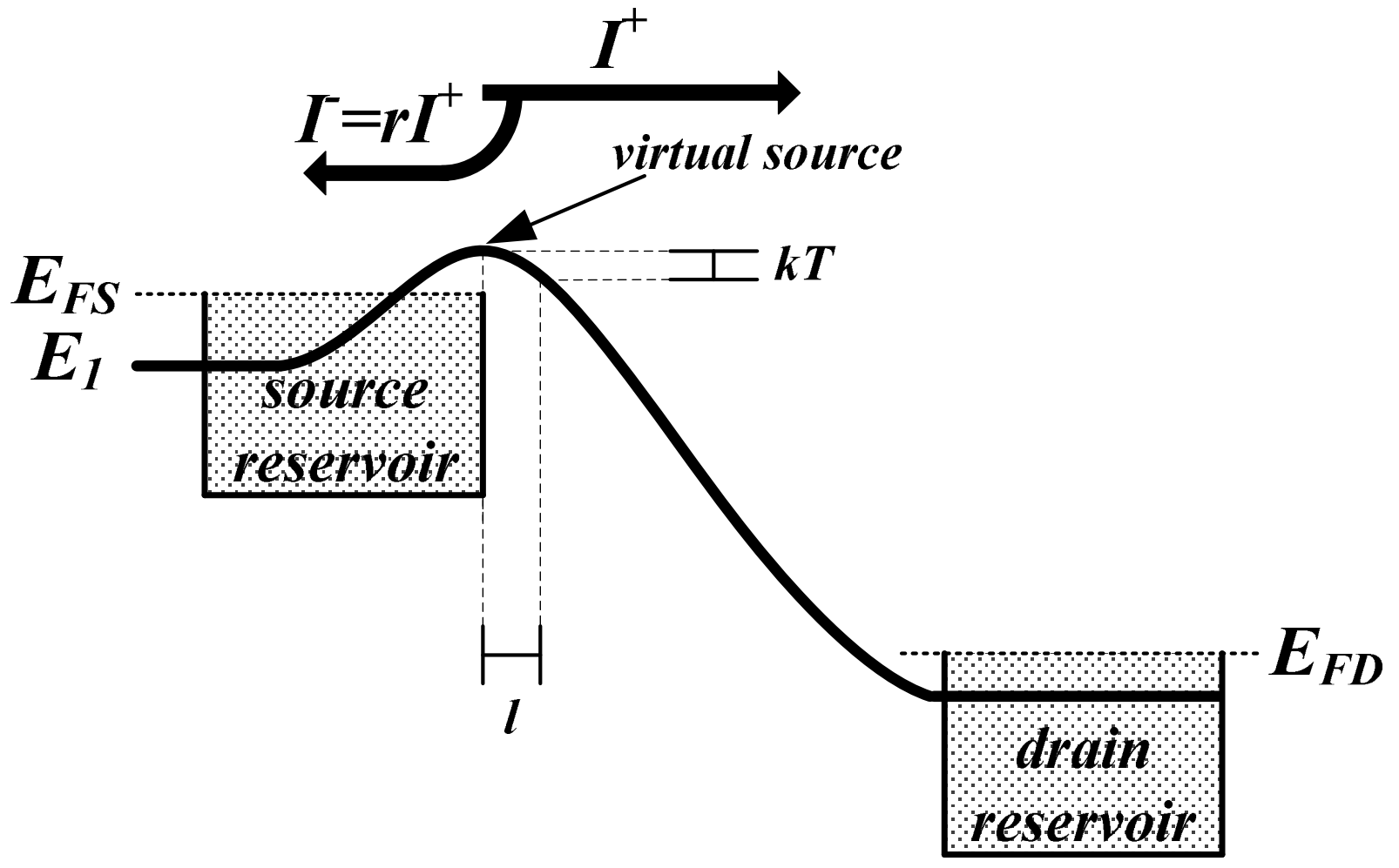

Figure 1

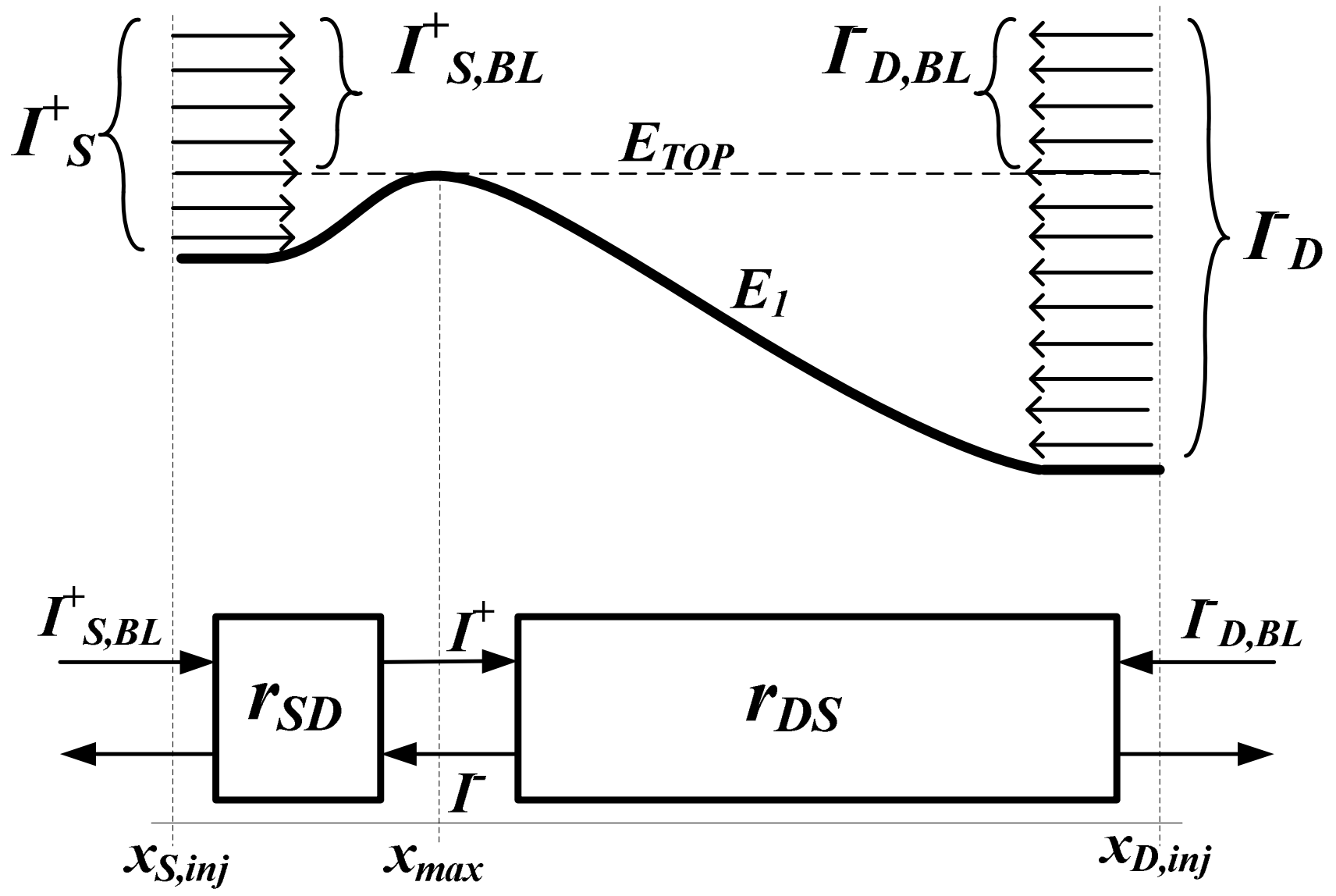

Figure 2

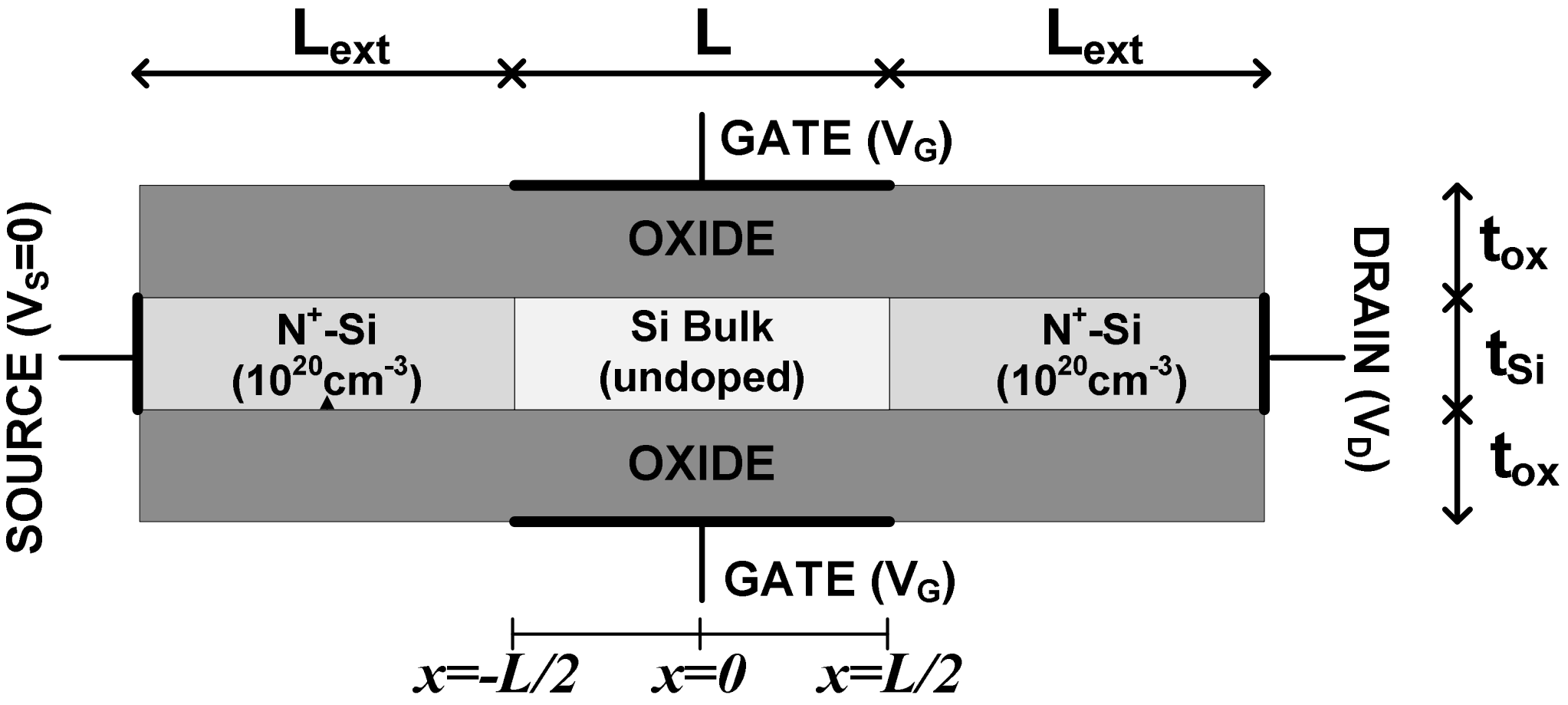

Figure 3

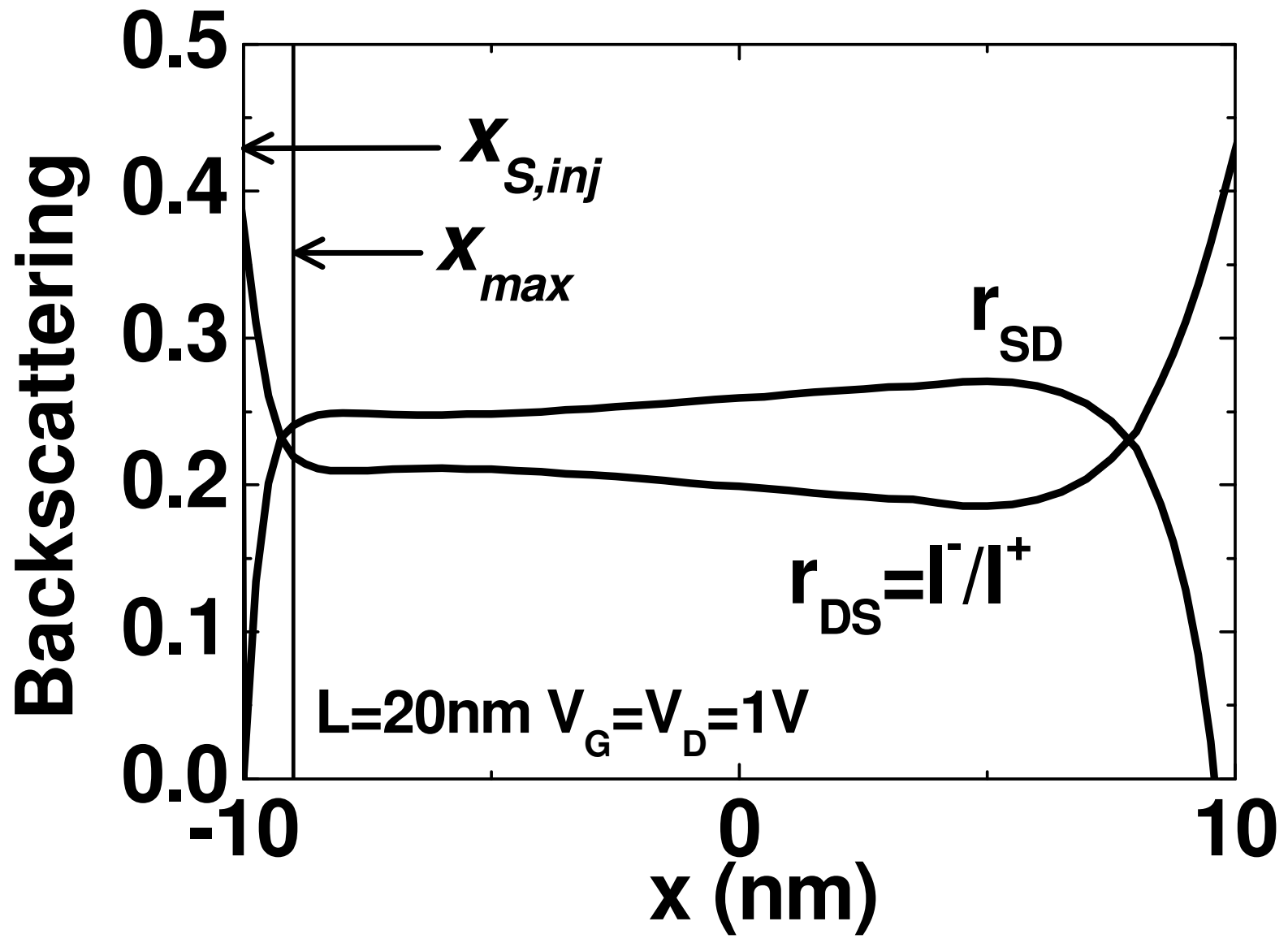

Figure 4

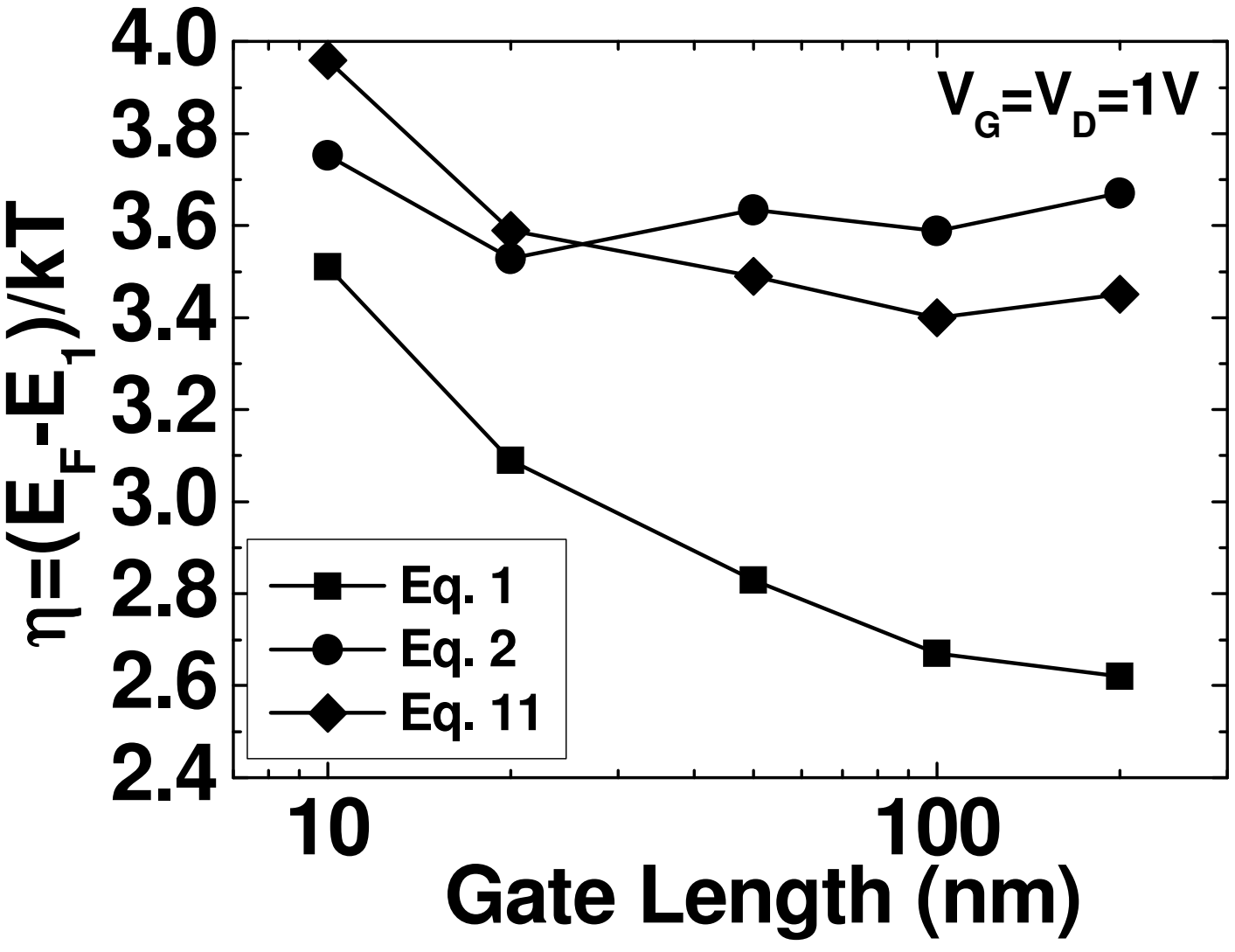

Figure 5

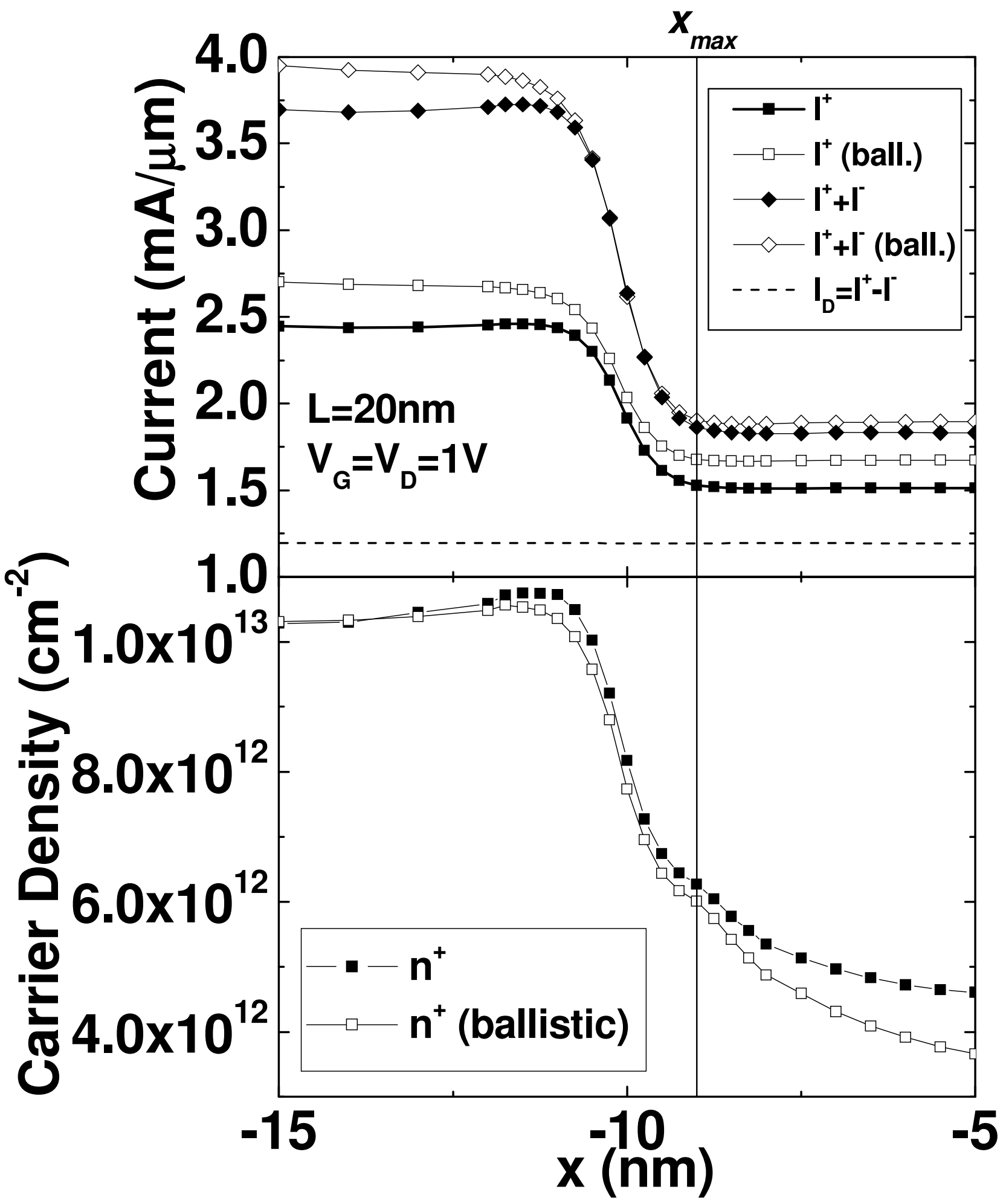

Figure 6

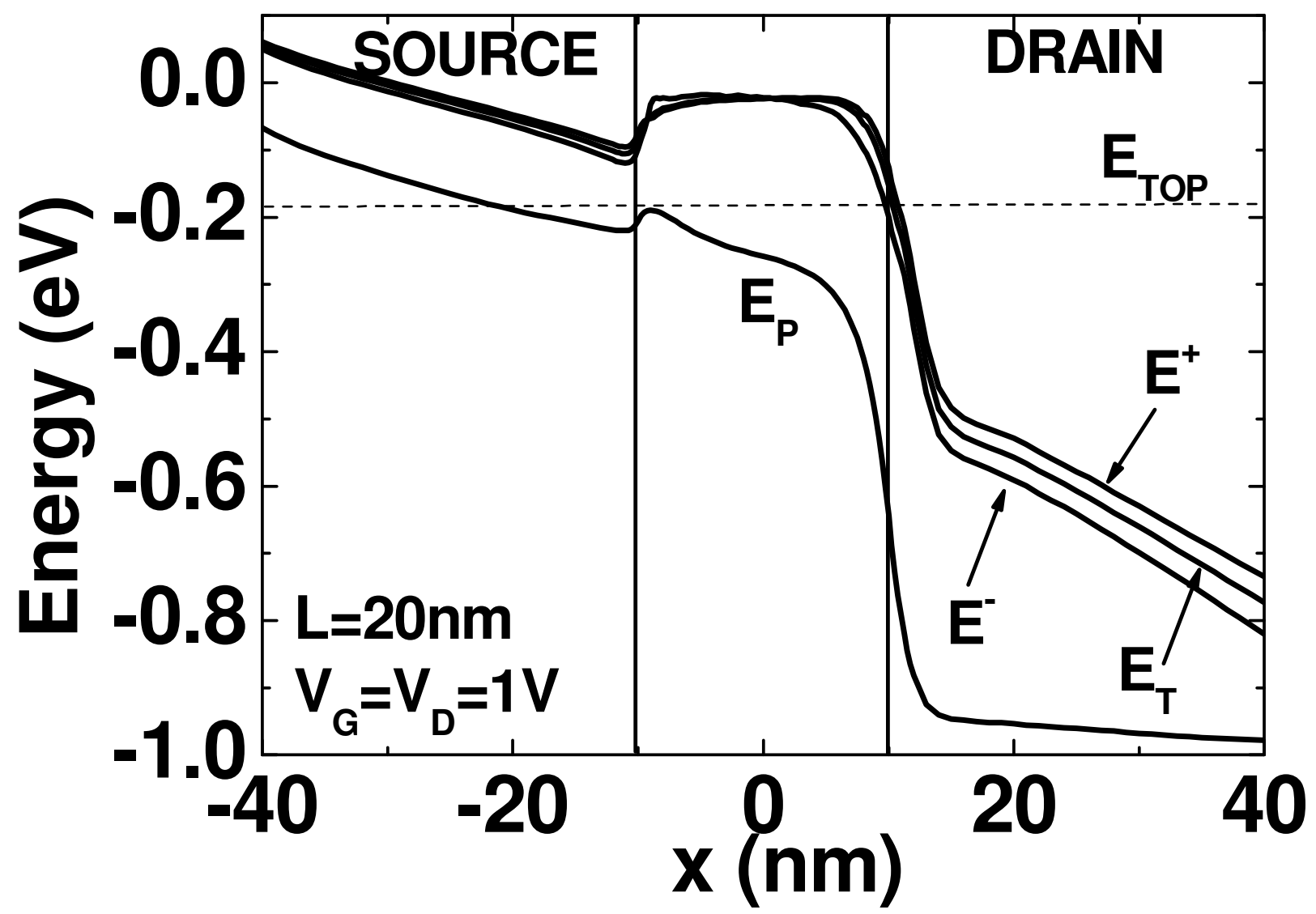

Figure 7

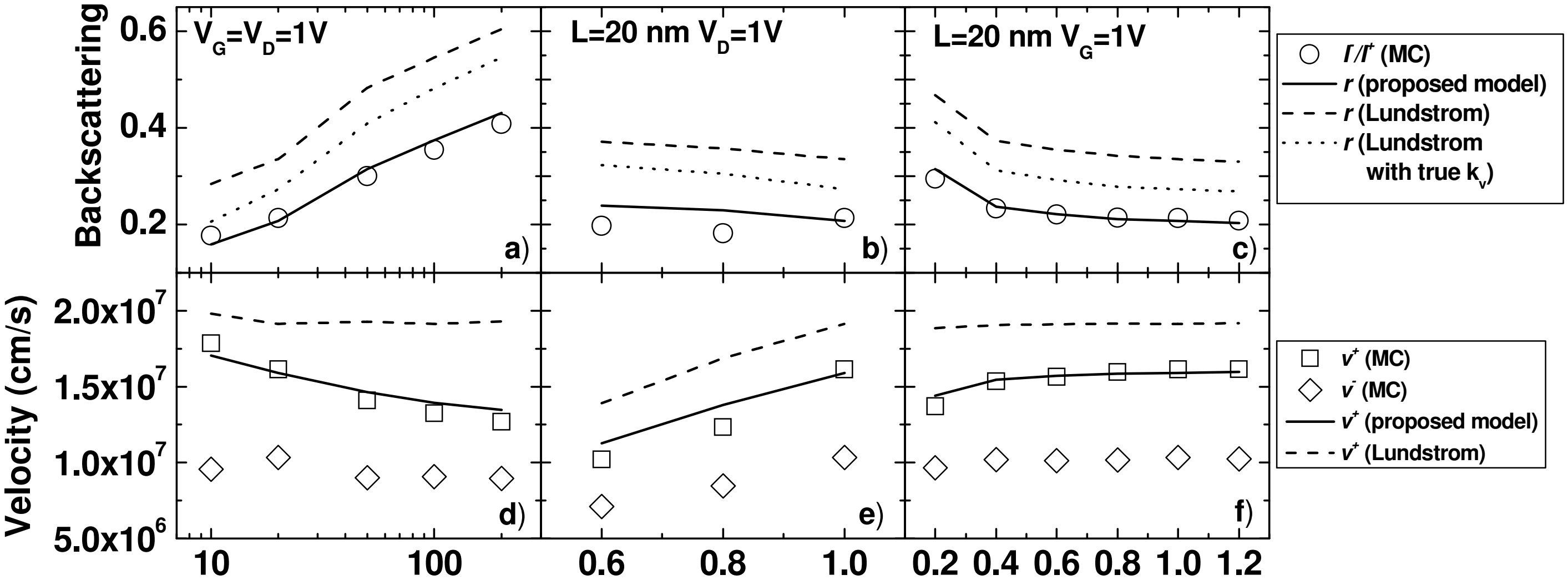

Figure 8

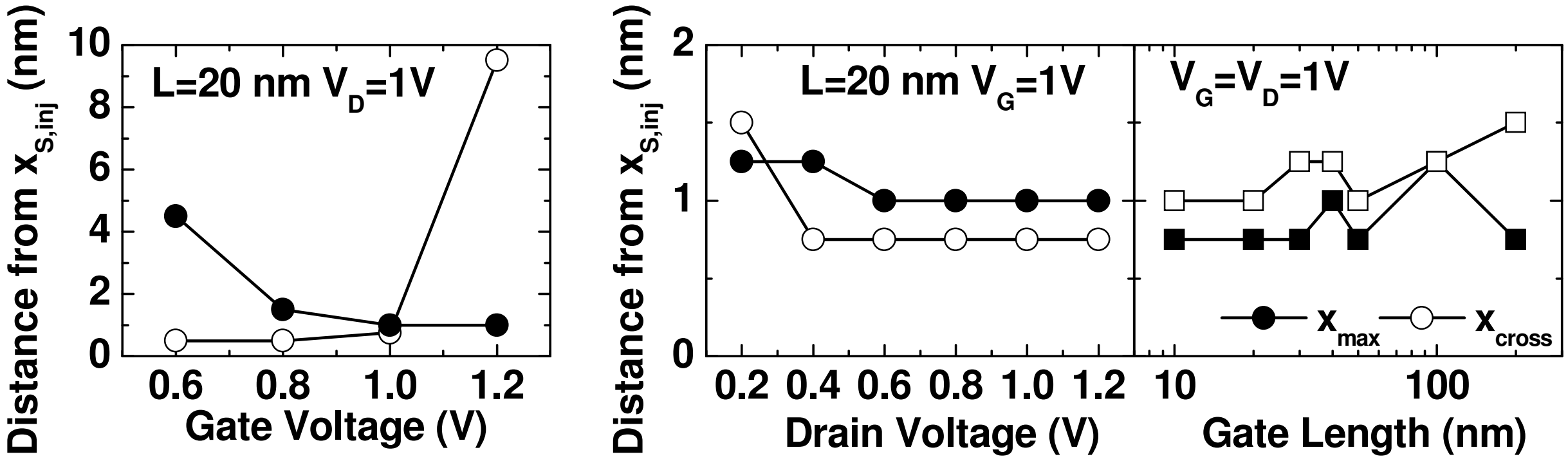

Figure 9

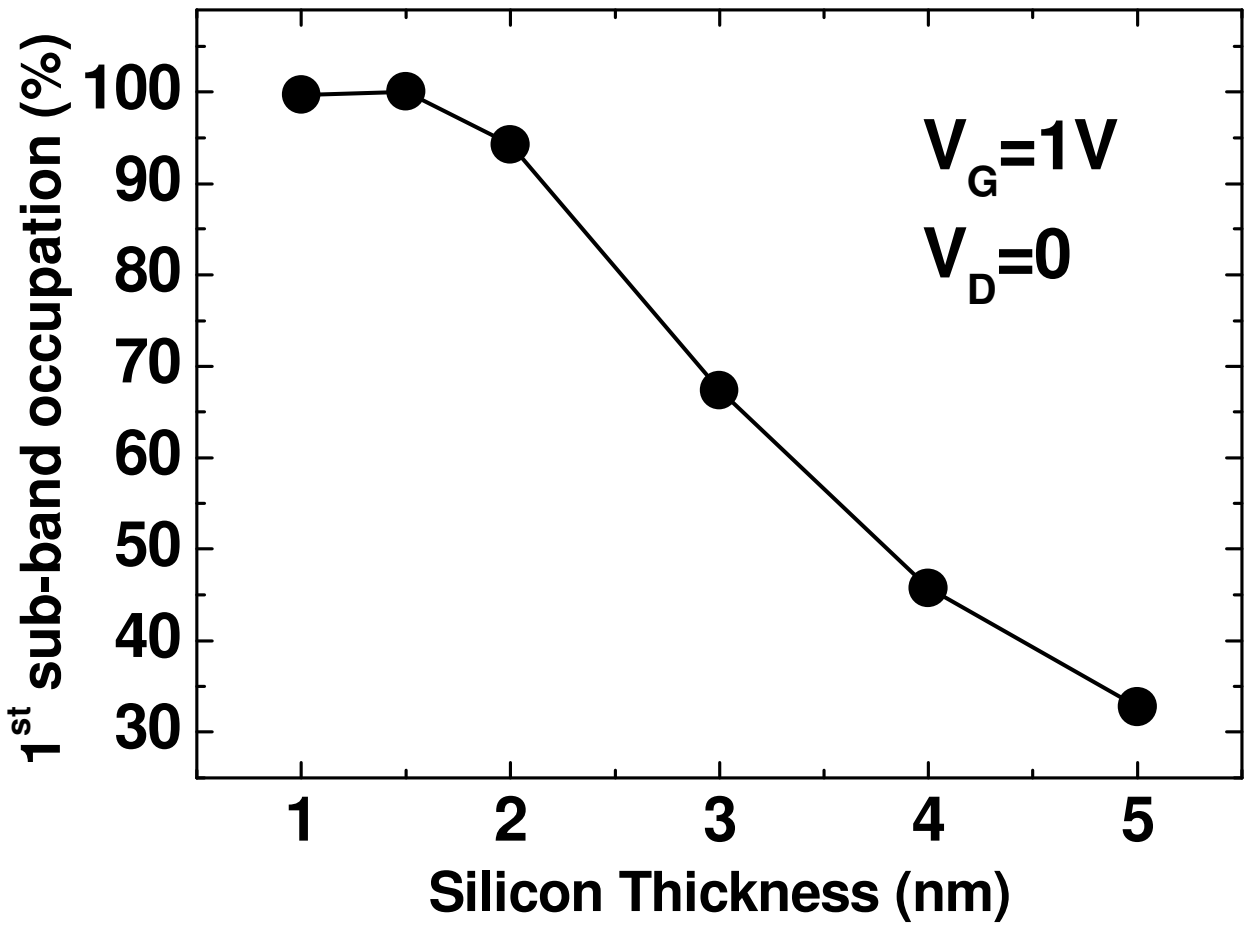

Figure 10